\documentclass[aps,preprint,amssymb,showpacs]{revtex4}
\usepackage{graphicx}
\usepackage{psfrag}
\newcommand{\be}{\begin{eqnarray}}
\newcommand{\ee}{\end{eqnarray}}
\newcommand{\up}{\uparrow}
\newcommand{\down}{\downarrow}
\newcommand{\ket}[1]{\vert\,{#1}\rangle}
\newcommand{\eps}{\epsilon}
\newcommand{\al}{\alpha}

\begin{document}
\title{Chiral odd GPDs in transverse and longitudinal impact 
parameter spaces}
\author{\bf D. Chakrabarti$^a$, R. Manohar$^b$, A. Mukherjee$^b$}
\affiliation{$^a$ Department of Physics, 
Swansea University, Singleton Park, Swansea, SA2 8PP, UK.\\
$^b$ Department of Physics,
Indian Institute of Technology, Powai, Mumbai 400076,
India.}
\date{\today}
\begin{abstract}
We investigate the chiral odd generalized parton distributions (GPDs)
for non-zero skewness $\zeta$ in transverse and longitudinal position 
spaces by taking Fourier transform with respect to the transverse and 
longitudinal momentum transfer respectively. We present overlap 
formulas for the chiral-odd GPDs in terms of light-front 
wave functions (LFWFs) of the proton both in the ERBL and DGLAP regions. 
We calculate them in a field theory inspired model of a
relativistic spin $1/2$ composite state with the correct correlation
between the different LFWFs in Fock space,
namely that of the quantum fluctuations of an electron in a 
generalized form of QED. We show the spin-orbit correlation effect of the
two-particle LFWF as well as the correlation between the constituent
spin and the transverse spin of the target.
\end{abstract}


\maketitle
\section{Introduction}
Generalized parton distributions (GPDs) give a unified picture of the
nucleon, in the sense that $x$ moments of them give the form factors
accessible in exclusive processes whereas in the forward limit they reduce to
parton distributions, accessible in inclusive processes (see \cite{rev} for
example). At zero skewness $ \zeta $, if one performs a Fourier
transform (FT) of the GPDs with respect to (wrt) the momentum
transfer in the transverse direction $\Delta_\perp$, one gets the
so called impact parameter dependent parton distributions, 
which tell us how the partons of a given longitudinal
momentum are distributed in transverse position (or impact
parameter $b_\perp$) space. These obey certain positivity
constraints and unlike the GPDs themselves, have probabilistic
interpretation \cite{bur}. As transverse boost on the light-front is
a Galilean boost, there are no relativistic corrections. Impact parameter 
dependent pdfs are 
defined for nucleon states localized in the transverse position space at
$R_\perp$. In order to avoid a singular normalization constant,
one can take a wave packet state. A wave packet state which is
transversely polarized is shifted sideways in the impact parameter
space \cite{burchi}. An interesting interpretation of
Ji's angular momentum sum rule \cite {ji} is obtained in terms of the
impact parameter dependent pdfs \cite{burchi}. On the other hand, in
\cite{hadron_optics}, real and imaginary parts of the DVCS amplitudes are
expressed in longitudinal position space by introducing a longitudinal
impact parameter $\sigma$ conjugate to the skewness $\zeta$, and it was shown that
the DVCS amplitude show certain diffraction pattern in the longitudinal
position space. Since Lorentz boosts are
kinematical in the front form, the correlation determined in the
three-dimensional $b_\perp, \sigma$ space is frame-independent.
As GPDs depend on a sharp $x$, the Heisenberg uncertainty
relation restricts the longitudinal position space
interpretation of GPDs themselves. It has, however, been shown 
in \cite{wigner} that one can define a quantum mechanical Wigner 
distribution for the relativistic quarks and gluons inside the proton. 
Integrating over $k^-$ and $k^\perp$, one
obtains a four dimensional quantum distribution which is a function
of ${\vec{r}}$ and $k^+$ where  ${\vec{r}}$ is the quark 
position vector defined in the rest frame of the proton. These
distributions are related to the FT of GPDs in the same frame. This gives a
3D position space picture of the GPDs and of the proton, within the
limitations mentioned above.

At leading twist, there are three forward parton distributions
(pdfs), namely, the unpolarized, helicity and transversity
distributions. Similarly, three leading twist generalized quark
distributions can be defined which in the forward limit, reduce to
these three forward pdfs. The third one is chiral odd and is
called the generalized transversity distribution $F_T$. This is defined as
the off-forward matrix element of the bilocal tensor charge operator.
It is parametrized in terms of four GPDs, namely $H_T$, $\tilde H_T$,
$E_T$ and $\tilde E_T$ in  the most general way \cite{markus,chiral,burchi}.
Unlike $E$, which gives a sideways shift in the unpolarized quark 
density in a transversely polarized nucleon, the chiral-odd GPDs affect the
transversely polarized quark distribution both in unpolarized and
in transversely polarized nucleon in various ways. A relation for the
transverse total angular momentum of the quarks has been proposed 
in \cite{burchi}, in analogy with Ji's relation, which involves a 
combination of second moments of $H_T, E_T$ and $\tilde H_T$ in the forward
limit. $\tilde E_T$ does not contribute when skewness $\zeta=0$, as it 
is an odd function of $\zeta$. $H_T$ reduces to the transversity distribution
in the forward limit when the momentum transfer is zero. Unlike
the chiral even GPDs, information about which can be and has been
obtained from deeply virtual Compton scattering and hard exclusive
meson production, it is very difficult to measure the chiral odd
GPDs. That is because, being chiral odd, they have to combine with another
chiral odd object in the amplitude. In \cite{double}, a proposal to
measure $H_T$ has been given in photo or electroproduction
of a longitudinally polarized vector meson
$\rho^0$ via two gluon  fusion; this meson is separated by a large rapidity
gap from other transversely polarized $\rho^+$ and the scattered neutron.
The scattering amplitude factorizes and involves $H_T(x,\zeta,0)$ at zero
momentum transfer as well as the chiral odd light-cone distribution
amplitude for the transversely polarized meson. In \cite{gary}, the 
exclusive process $\gamma^* P \rightarrow \pi^0 P$ has been suggested 
to measure the tensor charge. However, one has to look at the helicity flip 
part which is a higher twist contribution.  There
is also  a prospect of gaining information about the Mellin
moments of chiral odd GPDs from lattice QCD \cite{lattice}. They have been
investigated in several models, the first being the bag model
\cite{scopetta}, where only $H_T$ has been found to be non-zero.
In \cite{barbara}, they have been calculated in a constituent quark model,
a model  independent overlap in the DGLAP region is also given.
$H_T(x,\zeta,0)$ is also modeled in the ERBL region $x < \zeta$ in
\cite{double}. The possibility of getting model independent relations between
transverse momentum dependent parton distributions and impact parameter 
representation of GPDs at $\zeta=0$ has been investigated in \cite{metz},
however it was concluded that such relations are model dependent. In a
previous work we have investigated the chiral odd GPDs for a simple
spin-$1/2$ composite particle for $\zeta=0$ in impact parameter space
\cite{harleen}.

In this work, we present overlap
formulas for the chiral odd GPDs in the terms of the LFWFs both in the
DGLAP ($n \rightarrow n$) and ERBL ( $n+1 \rightarrow n-1$) regions.
We investigate them in a simple model, namely  for
the quantum fluctuations  of a  lepton in QED at
one-loop order \cite{drell}, the same system which gives the Schwinger
anomalous moment $\alpha/2 \pi$. We generalize this analysis
by assigning a mass $M$ to the external electrons and a different
mass $m$ to the internal electron lines and a mass $\lambda$ to the
internal photon lines with $M < m + \lambda$ for stability. 
In effect, we shall represent a spin-${1\over 2}$ system as a composite 
of a spin-${1\over 2}$ fermion and a spin-$1$ vector boson 
\cite{hadron_optics, dis,dip,dip2,marc}. This field theory inspired model 
has the correct correlation between the Fock components of the state 
as governed by the light-front
eigenvalue equation, something that is extremely difficult to achieve in
phenomenological models. Also, it gives an intuitive
understanding of the spin and orbital angular momentum of a
composite relativistic system \cite{orbit}. GPDs in this model satisfy
general properties like polynomiality and positivity. So it is interesting
to investigate the  properties of GPDs in this model. 
By taking Fourier transform (FT) with respect to $\Delta_\perp$, we express the
GPDs in transverse position space and by taking a FT with respect to $\zeta$
we expressed them in longitudinal position space.  

\section{Overlap Representation}
The chiral odd GPDs are expressed as the off forward matrix element of the
bilocal tensor charge operator on the light cone. These involve a
helicity flip of the quark.

We use the parametrization of \cite{burchi} for the chiral odd GPDs:
\be
F^{T \lambda',\lambda}_j &=& P^+ \int {dz^-\over 2 \pi} e^{i P^+z^-\over 2} \langle P', \lambda' \mid
{\bar \psi} ( {- z^-\over 2})\sigma^{+j} \gamma_5 \psi ({z^- \over 2}) \mid P, 
\lambda \rangle
\nonumber\\&& = H_T(x,\zeta,t) {\bar u} (P') \sigma^{+j} \gamma_5 u(P) 
+ \tilde H_T (x,\zeta,t) \eps^{+j \al \beta} {\bar u}(P') {\Delta_\al 
P_\beta \over
M^2} u(P)\nonumber\\&&+ E_T(x, \xi,t)\eps^{+j \al \beta} {\bar u}(P') 
{\Delta_\al \gamma_\beta \over
2 M} u(P)+\tilde E_T(x,\zeta,t) \eps^{+j \al \beta} {\bar u}(P') {P_\al 
\gamma_\beta \over M} u(P).\label{GPD}
\ee        

We choose the frame where the initial and final momenta of the proton with 
mass $M$ are:
\be
P&=&\left( P^+, 0_\perp, {M^2\over P^+ }\right),\\
P'&=&\left( (1-\zeta)P^+, -\Delta_\perp, {M^2+\Delta_\perp^2\over 
(1-\zeta)P^+ } \right).
\ee
So, the momentum transfered from the target is
\be
\Delta=P-P'=\left( \zeta P^+, \Delta_\perp, {t+\Delta_\perp^2 \over \zeta P^+} 
\right),
\ee
where $t=\Delta^2$.
Following \cite{overlap} we expand the proton state of momentum  $P$ and
helicity $\lambda$ in terms of
multi-particle light-front wave functions:
\be
\mid P, \lambda \rangle &=& \sum_{n} \prod_{i=1}^n {dx_i d^2 k_{\perp i} \over
\sqrt{x_i} 16 \pi^3} 16 \pi^3 \delta(1- \sum_{i=1}^n x_i) \delta^2
(\sum_{i=1}^n k_{\perp i})\nonumber\\&&~~\psi_n(x_i, k_{\perp i}, \lambda_i)
\mid n, x_i P^+, x_i P_\perp + k_{\perp i}, \lambda_i \rangle;
\ee
here $x_i=k_i^+/P^+$  is the light cone momentum fraction and $k_{\perp i}$ 
represent the relative transverse momentum of the $i$ th constituent. The
physical transverse momenta are $p_{\perp i} = x_i P_\perp + k_{\perp i}$.
$\lambda_i$ are the light-cone helicities. The light front wave functions 
$\psi_n$  are independent of $P^+$ and $P_\perp$
and are boost invariant.

Like the chiral even GPDs, here too there are diagonal $n \to n$ overlaps in
the kinematical region $\zeta < x < 1$ and $\zeta-1 < x < 0$. In the region 
$0< x <\zeta$ there are off diagonal $n+1 \to n-1$ overlaps.  
The overlap representation of the chiral odd GPDs in terms of light front 
wave functions is given by :
\be
F_{1,n \rightarrow n}^{T \lambda', \lambda}&=& 
(1-\zeta)^{1-{n\over 2}} \sum_{n,\lambda_i} \int \Pi_{i=1}^n {dx_i d^2
k_\perp^i\over 16 \pi^3} 16 \pi^3 \delta(1-\sum_j x_j)
\delta^2(\sum_{j=1}^n k_\perp^j) \delta(x-x_1) \nonumber\\&& \psi_n^{\lambda'*}(x_i',
{k'}_\perp^i,\lambda'_i) \psi_n^{\lambda}(x_i,
k_\perp^i,\lambda_i)                
\delta_{\lambda_1',-\lambda_1}[ \delta_{\lambda_i',\lambda_i}(i=2,..n)];    
\ee
where ${x'}_i={x_i\over 1-\zeta}; ~{k'}_\perp^i=k_\perp^i+{x_i\over
1-\zeta} \Delta_\perp$ for $i=2,....n$ and  ${x'}_1={x_1-\zeta \over 1-\zeta}; 
~{k'}_\perp^1=k_\perp^1-{1-x_1\over
1-\zeta} \Delta_\perp$.

\be
F_{1, n+1 \rightarrow n-1}^{T \lambda', \lambda}&=&
{(1-\zeta)}^{3/2-n/2} \sum_{n,\lambda_i} \int \Pi_{i=1}^{n+1} {dx_i d^2
k_\perp^i\over 16 \pi^3} (16 \pi^3)^2 \delta(1-\sum_{j=1}^{n+1}
x_j) \delta^2(\sum_{j=1}^{n+1} k_\perp^i) \delta(x_{n+1}+x_1-\zeta)
\nonumber\\&&\delta^2
(k_{\perp n+1} + k_{1 \perp} -\Delta_\perp) \delta(x-x_1) 
\psi_{n-1}^{\lambda'*}(x_i',
{k'}_\perp^i,\lambda'_i) \psi_{n+1}^{\lambda}(x_i,
k_\perp^i,\lambda_i) \nonumber\\&&~~~~~~~~~~~~~~~~                
\delta_{\lambda_1',-\lambda_{n+1}} 
[\delta_{\lambda_i',\lambda_i}(i=2,..n)].    
\ee
where ${x'}_i={x_i\over 1-\zeta},~~~~~~~{k'}_\perp^i=k_\perp^i+{x_i\over
1-\zeta} \Delta_\perp$, for $i=2,...n$ label the $n-1$ spectators.
The overlaps are different from the chiral even GPDs as there is a helicity
flip of the quark. The above  overlap formulas can be used in any model
calculation of the chiral odd GPDs using LFWFs.  
\section{Chiral Odd GPDs in QED at One Loop}
Following \cite{drell, hadron_optics}, we take a simple composite spin
$1/2$ state, namely an electron in QED at one loop to investigate the GPDs. 
The light-front Fock state wavefunctions corresponding to the
quantum fluctuations of a physical electron can be systematically
evaluated in QED perturbation theory. The state is expanded in Fock 
space and there
are contributions from $\ket{e^- \gamma}$ and $\ket{e^- e^- e^+}$,
in addition to renormalizing the one-electron state. The
two-particle state is expanded as,
\begin{eqnarray}
\lefteqn{
\left|\Psi^{\uparrow}_{\rm two \ particle}(P^+, \vec P_\perp = \vec
0_\perp)\right> =
\int\frac{{\mathrm d} x \, {\mathrm d}^2
           {\vec k}_{\perp} }{\sqrt{x(1-x)}\, 16 \pi^3}
}
\label{vsn1}\\
&&
\left[ \ \ \,
\psi^{\uparrow}_{+\frac{1}{2}\, +1}(x,{\vec k}_{\perp})\,
\left| +\frac{1}{2}\, +1\, ;\,\, xP^+\, ,\,\, {\vec k}_{\perp}\right>
+\psi^{\uparrow}_{+\frac{1}{2}\, -1}(x,{\vec k}_{\perp})\,
\left| +\frac{1}{2}\, -1\, ;\,\, xP^+\, ,\,\, {\vec k}_{\perp}\right>
\right.
\nonumber\\
&&\left. {}
\psi^{\uparrow}_{+\frac{1}{2}\, +1}(x,{\vec k}_{\perp})\,
\left| +\frac{1}{2}\, +1\, ;\,\, xP^+\, ,\,\, {\vec k}_{\perp}\right>
+\psi^{\uparrow}_{+\frac{1}{2}\, -1}(x,{\vec k}_{\perp})\,
\left| +\frac{1}{2}\, -1\, ;\,\, xP^+\, ,\,\, {\vec k}_{\perp}\right>
\right.
\nonumber\\
&&\left. {}
+\psi^{\uparrow}_{-\frac{1}{2}\, +1} (x,{\vec k}_{\perp})\,
\left| -\frac{1}{2}\, +1\, ;\,\, xP^+\, ,\,\, {\vec k}_{\perp}\right>
+\psi^{\uparrow}_{-\frac{1}{2}\, -1} (x,{\vec k}_{\perp})\,
\left| -\frac{1}{2}\, -1\, ;\,\, xP^+\, ,\,\, {\vec k}_{\perp}\right>\
\right] \ ,
\nonumber
\end{eqnarray}
where the two-particle states $|s_{\rm f}^z, s_{\rm b}^z; \ x, {\vec
k}_{\perp} \rangle$ are normalized as in \cite{overlap}. $s_{\rm
f}^z$ and $s_{\rm b}^z$ denote the $z$-component of the spins of the
constituent fermion and boson, respectively, and the variables $x$
and ${\vec k}_{\perp}$ refer to the momentum of the fermion. The
light cone momentum fraction $x_i= {k_i^+\over P^+}$  satisfy $0<
x_i \le 1$ and $\sum_i x_i =1$. We employ the light-cone gauge $A^+=0$,
so that the gauge boson polarizations are physical. The
three-particle state has a similar expansion. Both the two- and
three-particle Fock state components are given in \cite{overlap}.
We here give the two-particle wave function for spin-up electron
\cite{orbit,drell,overlap}
\begin{equation}
\left
\{ \begin{array}{l}
\psi^{\uparrow}_{+\frac{1}{2}\, +1} (x,{\vec k}_{\perp})=-{\sqrt{2}}
\ \frac{-k^1+{i} k^2}{x(1-x)}\,
\varphi \ ,\\
\psi^{\uparrow}_{+\frac{1}{2}\, -1} (x,{\vec k}_{\perp})=-{\sqrt{2}}
\ \frac{k^1+{i} k^2}{1-x }\,
\varphi \ ,\\
\psi^{\uparrow}_{-\frac{1}{2}\, +1} (x,{\vec k}_{\perp})=-{\sqrt{2}}
\ (M-{m\over x})\,
\varphi \ ,\\
\psi^{\uparrow}_{-\frac{1}{2}\, -1} (x,{\vec k}_{\perp})=0\ ,
\end{array}
\right.
\label{vsn2}
\end{equation}
\begin{equation}
\varphi (x,{\vec k}_{\perp}) = \frac{e}{\sqrt{1-x}}\
\frac{1}{M^2-{{\vec k}_{\perp}^2+m^2 \over x}
-{{\vec k}_{\perp}^2+\lambda^2 \over 1-x}}\ .
\label{wfdenom}
\end{equation}
Similarly, the wave function for an electron with negative helicity
can also be obtained.

Following the same references, we work in a generalized form of QED
by assigning a mass $M$ to the external electrons and a different
mass $m$ to the internal electron lines and a mass $\lambda$ to the
internal photon lines. The idea behind this is to model the
structure of a composite fermion state with mass $M$ by a fermion
and a vector ``diquark" constituent with respective masses $m$ and
$\lambda$. The electron in QED also has a one-particle component
\begin{equation}
\left|\Psi_{\rm one \ particle}^{\uparrow , \downarrow}
(P^+, \vec P_\perp = \vec 0_\perp)\right> =
\int {{\mathrm d}x\, {\mathrm d}^2 {\vec{k}}_{\perp} \over
\sqrt{x}\, 16\pi^3}\ 16\pi^3 \delta (1-x)\,
{\delta}^2({\vec{k}}_{\perp})\
\psi_{(1)}\ \left| \pm {1\over 2} \, ;
xP^+, {\vec{k}}_{\perp} \right>
\label{bare1}
\end{equation}
where the one-constituent wavefunction is given by
\begin{equation}
\psi_{(1)} = \sqrt{Z} .
\label{oneparticle}
\end{equation}
Here $\sqrt Z$ is the wavefunction renormalization of the
one-particle state and ensures overall probability conservation.
Also, in order to regulate the ultraviolet
divergences we use a cutoff on the transverse momentum $k^\perp$.
If, instead of imposing a cutoff
on transverse momentum, we imposed a cutoff on the invariant
mass \cite{orbit}, then the divergences at $x=1$ would have been
regulated by the non-zero photon mass.

In the domain $\zeta <x <1$, there are diagonal $2 \to 2$ overlaps.
These correspond to (setting $j=1$ in Eqn.(\ref{GPD})) the helicity 
non-flip $F_1^{T \up \up}, 
F_1^{T \down \down}$ and helicity flip
contributions,  $F_1^{T \up\down},  F_1^{T \down \up}$, respectively, 
which can be written as,
\be
F_1^{T \up \up}= \int {d^2 k_\perp\over 16 \pi^3} \Big [ \psi^{* 1/2}_{1/2+1}
(x',k'_\perp) \psi^{1/2}_{-1/2+1} (x,k_\perp)+\psi^{* 1/2}_{-1/2+1} 
(x',k'_\perp) \psi^{1/2}_{1/2+1} (x,k_\perp) \Big ];\label{Fuu} 
\ee
\be
F_1^{T \up \down}= \int {d^2 k_\perp\over 16 \pi^3} \Big [ \psi^{* 1/2}_{1/2+1}
(x',k'_\perp) \psi^{-1/2}_{-1/2+1} (x,k_\perp)+\psi^{* 1/2}_{1/2-1} 
(x',k'_\perp) \psi^{-1/2}_{-1/2-1} (x,k_\perp) \Big ];\label{Fud} 
\ee
\be
F_1^{T \down \down}= \int {d^2 k_\perp\over 16 \pi^3} \Big [ \psi^{*
-1/2}_{1/2-1} (x',k'_\perp) \psi^{-1/2}_{-1/2-1} (x,k_\perp)+\psi^{*
-1/2}_{-1/2-1} (x',k'_\perp) \psi^{-1/2}_{1/2-1} (x,k_\perp) \Big ]; 
\label{Fdd}
\ee
\be
F_1^{T \down \up}= \int {d^2 k_\perp\over 16 \pi^3} \Big [ \psi^{*
-1/2}_{-1/2-1} (x',k'_\perp) \psi^{1/2}_{1/2-1} (x,k_\perp)+\psi^{*
-1/2}_{-1/2+1} (x',k'_\perp) \psi^{1/2}_{1/2+1} (x,k_\perp) \Big ]; 
\label{Fdu}\ee
where 
\be
x'={x-\zeta\over 1-\zeta}, ~~~~~~~~~~~~~~~~~k'_\perp=k_\perp-(1-x')
\Delta_\perp.
\ee
$F_1^{T \up \down}$ and $F_1^{T \down \up}$ receive contribution from the 
single particle sector of the Fock space, which is strictly at $x=1$
(wavefunction renormalization) \cite{dip2}. We exclude
$x=1$ by imposing a cutoff, and we do not consider this contribution
in this work.  It contributes also in the $3 \to 1 $ overlap, in the
ERBL region. However, 
the single particle contribution is important as it cancels the singularity
as $x \rightarrow 1$. This has been shown explicitly in the forward limit in
\cite{tran}, namely for the transversity distribution $h_1(x)$. The
coefficient of the logarithmic term in the expression of $h_1(x)$ gives the
correct splitting function for leading order evolution of $h_1(x)$; the
delta function from the single particle sector providing the necessary 
'plus' prescription. In the off forward case, the cancellation occurs 
similarly, as shown for $F(x, \xi,t)$ in \cite{marc}. The behavior at 
$x=0,1$ can be improved by differentiating the LFWFs with respect to 
$M^2$ \cite{har}. The $x \to 0$ limit can be improved as well by a
different method \cite{rad}.

The GPDs are zero in the domain $\zeta-1 < x < 0$, which
corresponds to emission and reabsorption of an $e^+$ from a physical
electron. Contributions to the GPDs in that domain only appear beyond 
one-loop level.

In our model $F_1^{T \up \down}= F_1^{T \down \up}$ and thus
\be
{F_1^{T \up \down}- F_1^{T \down \up}\over 2}
&=& \tilde H_T {1\over \sqrt{1-\zeta}}{\Delta^1\Delta^2\over M^2}\nonumber \\
&=& 0.
\ee 
Since the above equation is true for arbitrary $\Delta_1,~ \Delta_2$,  it 
implies $\tilde H_T =0$ in our model.
We define the combinations :
\be
(-i \Delta_2) A_1 &=& {F_1^{T \up \up}+ F_1^{T \down \down}\over 2}
\nonumber\\&=& {-i \Delta_2\over 2 M \sqrt{1-\zeta}} \Big [ (2-\zeta) E_T +
2 \tilde E_T  \Big ]; \label{A1}\\
(\Delta_1) A_2 &=& {F_1^{T \up \up}- F_1^{T \down \down}\over 2}
\nonumber\\&=& {\Delta_1\over 2 M \sqrt{-\zeta}} \Big [ \zeta E_T +
2 \tilde E_T  \Big ];  \label{A2}\\
 i A_3 &=&{F_1^{T \up \down}+ F_1^{T \down \up}\over 2}\nonumber \\
&=&i\Big [ 2 H_T \sqrt{1-\zeta} - {\zeta^2\over 2 \sqrt{1-\zeta}} E_T
-{\zeta\over \sqrt{1-\zeta}} \tilde E_T \Big ].  \label{A3}
\ee 
From the above equations we have 
\be
E_T(x,\zeta,t)&=&{M\over \sqrt{1-\zeta}}(A_1-A_2), \\
\tilde E_T(x,\zeta,t)&=&{M\over 2\sqrt{1-\zeta}}\{(2-\zeta)A_2-\zeta A_1\},\\
H_T(x,\zeta,t)&=&{1\over 2\sqrt{1-\zeta}}(A_3+ \zeta M A_2).
\ee
\begin{figure}[t]
\includegraphics[width=7cm,height=7cm,clip]{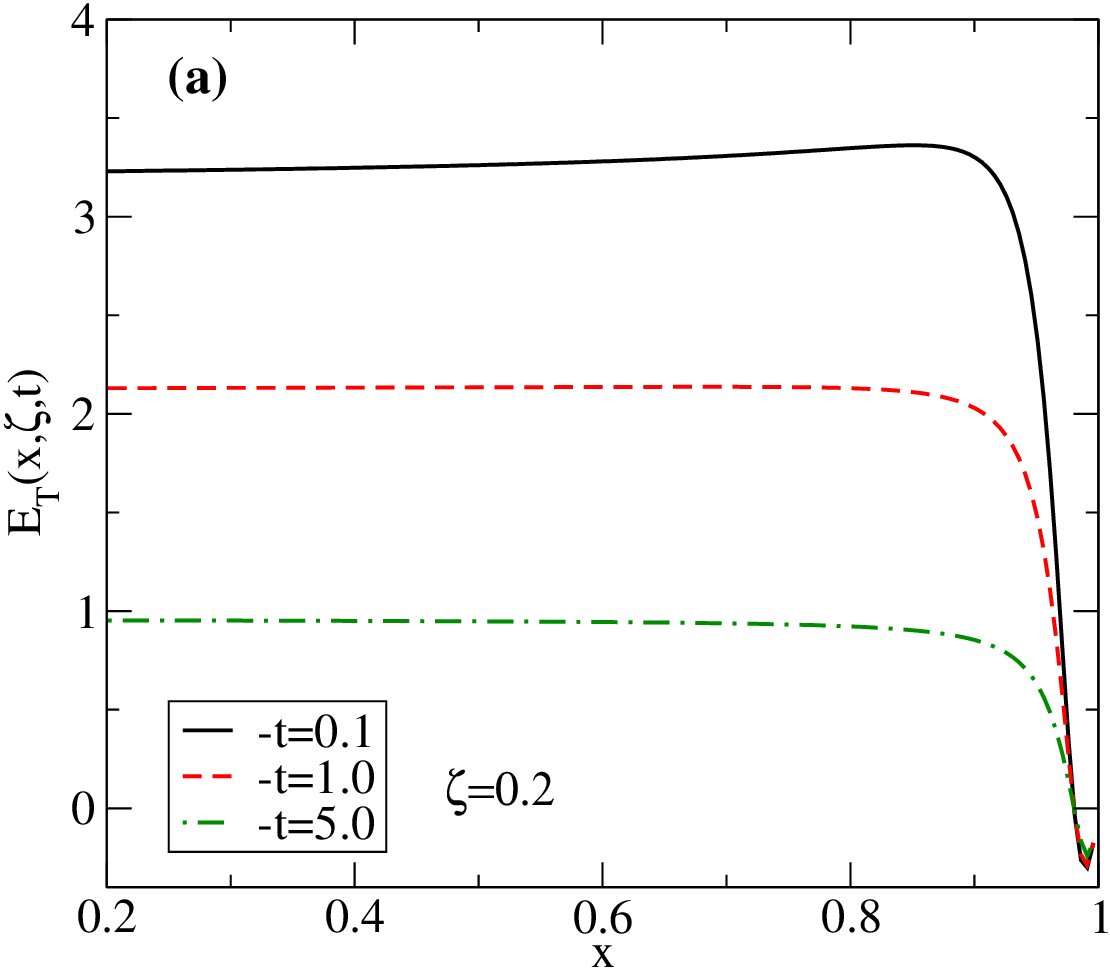}
\hspace{0.2cm}%
\includegraphics[width=7cm,height=7cm,clip]{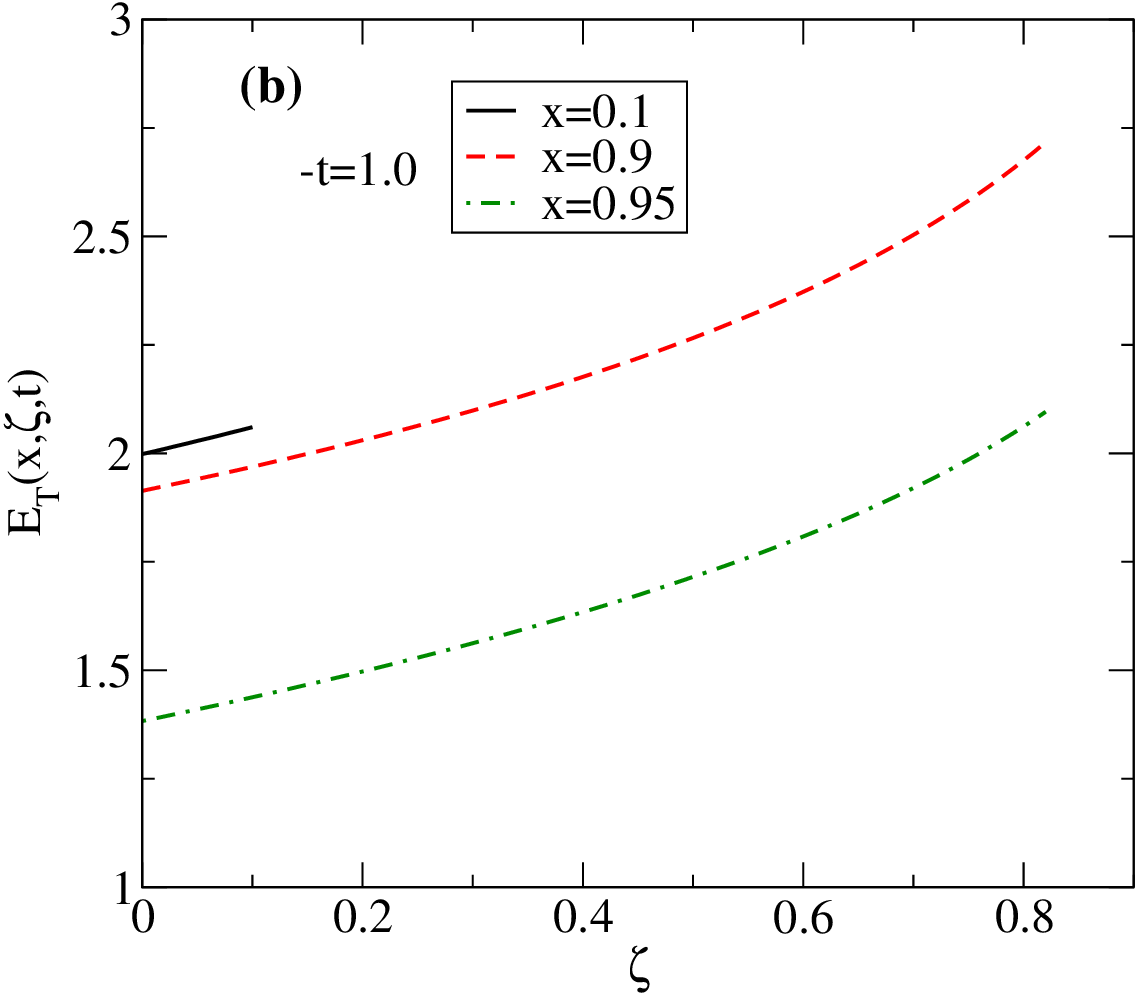}
{\caption{(Color online) Plot of $E_T(x,\zeta,t)$ (a) vs. $x$ for a fixed value of
$\zeta=0.2$ and different values of $t$ in ${\mathrm{MeV}}^2$; 
(b) vs. $\zeta$ for different values of $x$ and for fixed $-t=1~ 
{\mathrm{MeV}}^2$}}  
\end{figure}
Explicit matrix element calculation using Eqns(\ref{Fuu}-\ref{Fdu})
in our model gives
\be  
A_1&=&{\pi e^2\over 8 \pi^3 \sqrt{1-\zeta}}\Big [ \{(M-{m\over x}) x (1-\zeta)
\nonumber\\&&~~~~~~~~~~~
-(M-{m\over x'})x'\}(1-x') I_1 -(M-{m\over x})x (1-x) I_2 \Big ],\\ 
A_2&=&{\pi e^2\over 8 \pi^3 \sqrt{1-\zeta}} \Big \{(M-{m\over x})(1-x)x I_2 
-\Big [ (M-{m\over x}) (1-\zeta) x 
\nonumber\\&&~~~~~~~~~~~~~+(M-{m\over x'})x' \Big ](1-x')I_1  \Big \},\\   
A_3&=&{e^2\over 16 \pi^3}  {(x+x')\over \sqrt{(1-x)(1-x')}} \Big [ -I_3-I_4 
+B(x,\zeta) \pi I_2 \Big ];
\ee
here $B(x,\zeta)= M^2 x'(1-x')-m^2(1-x')-\lambda^2 x'+M^2 x (1-x) -m^2 (1-x) 
-\lambda^2 x-(1-x')^2 \Delta_\perp^2$.
\be
I_1 &=&\int_0^1 dy{1-y\over Q(y)},\\
I_2 &=&\int_0^1 dy{1\over Q(y)};
\ee
where $Q(y)=y(1-y)(1-x')^2 \Delta_\perp^2-y(M^2x(1-x)-m^2(1-x)-\lambda^2x)
-(1-y)(M^2x'(1-x')-m^2(1-x')-\lambda^2 x')$ and 
\begin{figure}[t]
\begin{center}
\includegraphics[clip,width=7.0cm]{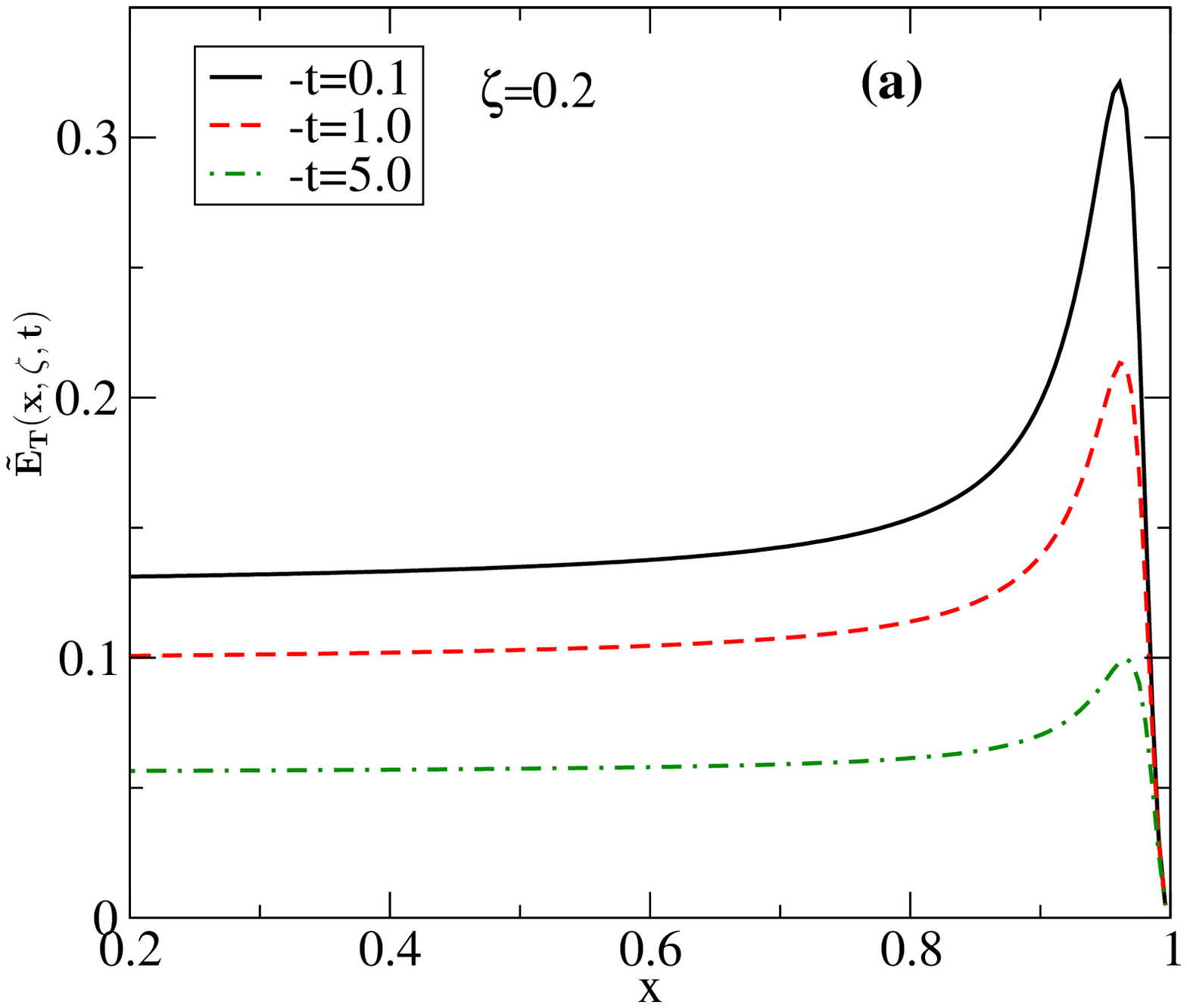}
\hspace{0.2cm}%
\includegraphics[clip,width=7.0cm]{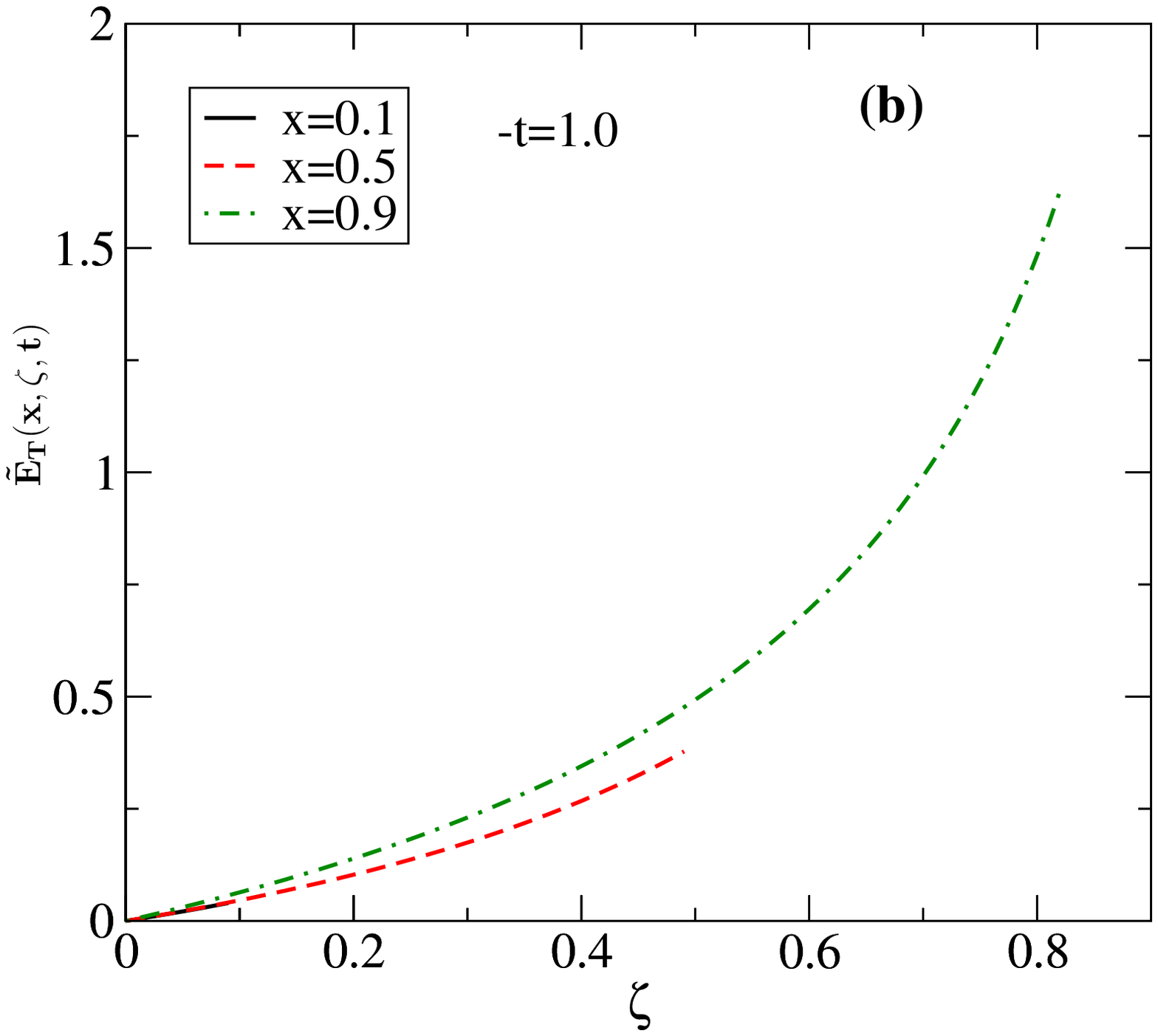}
\end{center}
{\caption{(Color online) Plot of $\tilde{E}_T(x,\zeta,t)$ (a) vs. $x$ for a fixed value of
$\zeta=0.2$ and different values of $t$ in ${\mathrm{MeV}}^2$; 
(b) vs. $\zeta$ for different values of $x$ and for fixed $-t=1~ 
{\mathrm{MeV}}^2$}}  
\end{figure}
\be
I_3&=&\int_0^\Lambda d^2k_\perp{1\over -k_\perp^2+M^2 x(1-x)-m^2(1-x)
-\lambda^2x} \nonumber\\
&=&-\pi\ln\mid{\Lambda^2 \over m^2(1-x)+\lambda^2x-M^2 x(1-x)}\mid;\\
I_4&=&\int_0^\Lambda d^2k_\perp{1\over -{k'_\perp}^2+M^2 x'(1-x')
-m^2(1-x')-\lambda^2x'} \nonumber\\
&=&-\pi\ln\mid{\Lambda^2 \over m^2(1-x')+\lambda^2x'-M^2 x'(1-x')}\mid.
\ee   

So, we have the expressions for the GPDs 
\be
E_T(x,\zeta,t)&=&-{e^2\over 8 \pi^3}{2 M\pi\over 1-\zeta}(M-{m\over x})x(1-x)
I_5,
\ee
\be   
\tilde E_T(x,\zeta,t)&=&{e^2\over 8\pi^3}{M\pi\over 1-\zeta}\Big 
[-(1-x)\Big\{
(M-{m\over x})x+(M-{m\over x'})x'\Big\}I_1\nonumber\\&&~~~~~~~~~
+(M-{m\over x})x(1-x)I_2\Big ],
\ee
\be
H_T(x,\zeta,t)&=&{e^2\over 8\pi^3}{\pi\over 2}\Big [{x+x'\over 2(1-x)} 
\ln({\Lambda^4 \over DD'})+\Big\{{x+x'\over 2(1-x)}B(x,\zeta)
+{\zeta M \over 1-\zeta}(M-{m\over x})x(1-x)\Big\}I_2 \nonumber\\
&&-{\zeta M\over 1-\zeta}\Big\{(M-{m\over x})x(1-\zeta)+
(M-{m\over x'})x'\Big\}(1-x')I_1\Big ],
\ee
with
$$I_5=\int_0^1 dy {y\over Q(y)},$$ 
 $D=M^2 x (1-x) -m^2 (1-x) -\lambda^2 x$ and $D'=M^2 x' (1-x') -m^2 
(1-x') -\lambda^2 x'$.

All the numerical plots are performed in units of ${e^2/(8\pi^3)}$. We took 
$M=0.51$~  MeV, $m=0.5$~ MeV and $\lambda=0.02$~ MeV. 
In Figs. 1-3 (a), we have  plotted the GPDs $ E_T,~\tilde E_T$ and $H_T, $ as
functions of x for fixed values of $\zeta = 0.2$ and a fixed value of $t$. 
$H_T$ has logarithmic divergence, similar to the transversity distribution
\cite{tran}. For numerical analysis, we have used a cutoff $\Lambda=10$ MeV 
($ \Lambda>>M $) on transverse momentum. In Fig.(\ref{Lambda_dep}), we 
show the 
cutoff 
dependence of $H_T$.
However, one has to incorporate  the
single particle contribution to get the correct $+$ distribution and also
to get a finite answer at $x=1$, as stated above. Both $E_T$ and $\tilde E_T$ 
are  independent of $x$ at small and medium $x$
and at $x \to 1$, they are independent of $t$ \cite{bur} and approach zero.
Note that our analytic expressions for $H_T$ and $E_T$ agree with
the quark model calculation of \cite{metz} without the color factors, in the
limit $\zeta=0$ and $M=m$ and $\lambda=0$. Fig. 1-3 (b) present the
$\zeta$ behaviour of the above GPDs for fixed $t=-1.0~ {\mathrm{MeV}}^2$ and
different values of $x$. Note that as we are plotting in the DGLAP region, $x
> \zeta$. Both $E_T$ and $\tilde E_T$ increases in 
magnitude with increase       
of $\zeta$ but $H_T$ decreases. $\tilde E_T$ is zero at $\zeta=0$.

\vspace{0.8cm}
\begin{figure}[t]
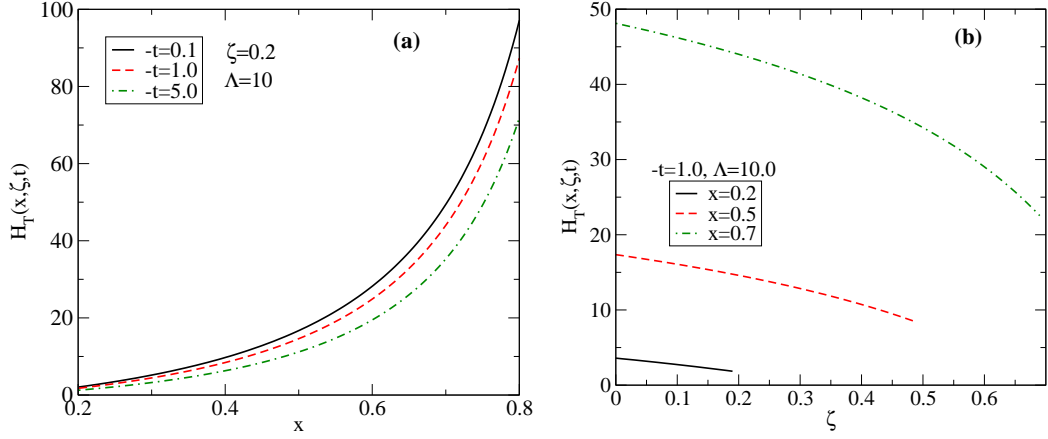

\begin{center}
\includegraphics[clip,width=7cm]{fig3a.eps}
\hspace{0.2cm}%
\includegraphics[clip,width=6.5cm]{fig3b.eps}
\end{center}
{\caption{(Color online) Plot of $H_T(x,\zeta,t)$ (a) vs. $x$ for a fixed value of
$\zeta=0.2$ and different values of $t$ in ${\mathrm{MeV}}^2$; 
(b) vs. $\zeta$ for different values of $x$ and for fixed $-t=1~ 
{\mathrm{MeV}}^2$. $\Lambda$ is in MeV.}}  
\end{figure}

\begin{figure}[ht]
\begin{center}
\includegraphics[clip,width=7cm]{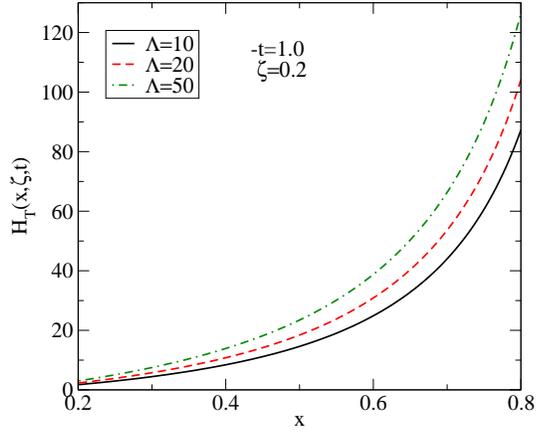}
\end{center}
{\caption{(Color online) $H_T(x,\zeta,t)$ for different UV cutoff $\Lambda$ 
(in MeV).}
\label{Lambda_dep} } 
\end{figure}

\section{GPDs in position  space}
\begin{figure}
\centering
\includegraphics[width=7cm,height=7cm,clip]{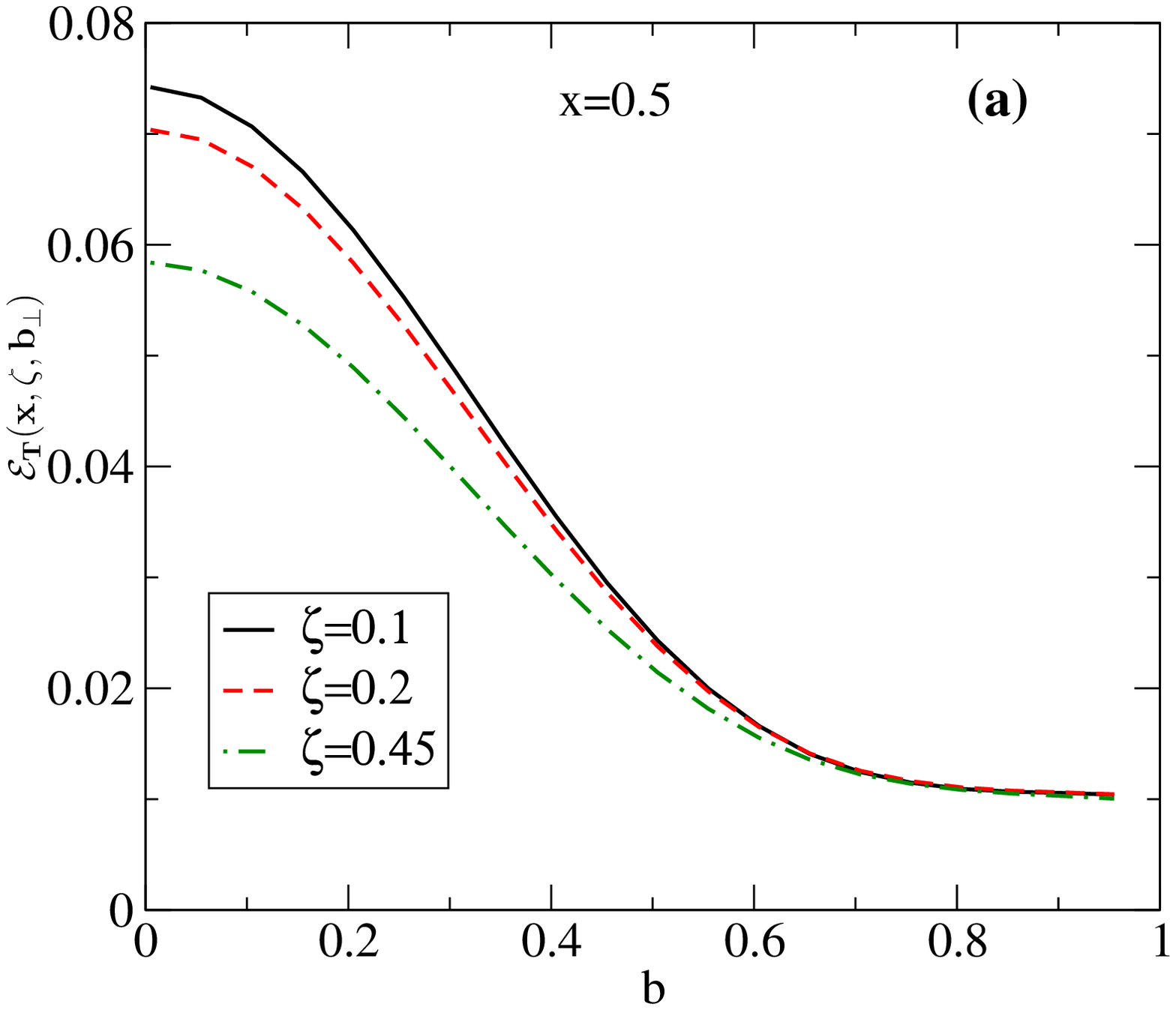}
\centering
\begin{minipage}[c]{0.9\textwidth}
\includegraphics[width=7cm,height=7cm,clip]{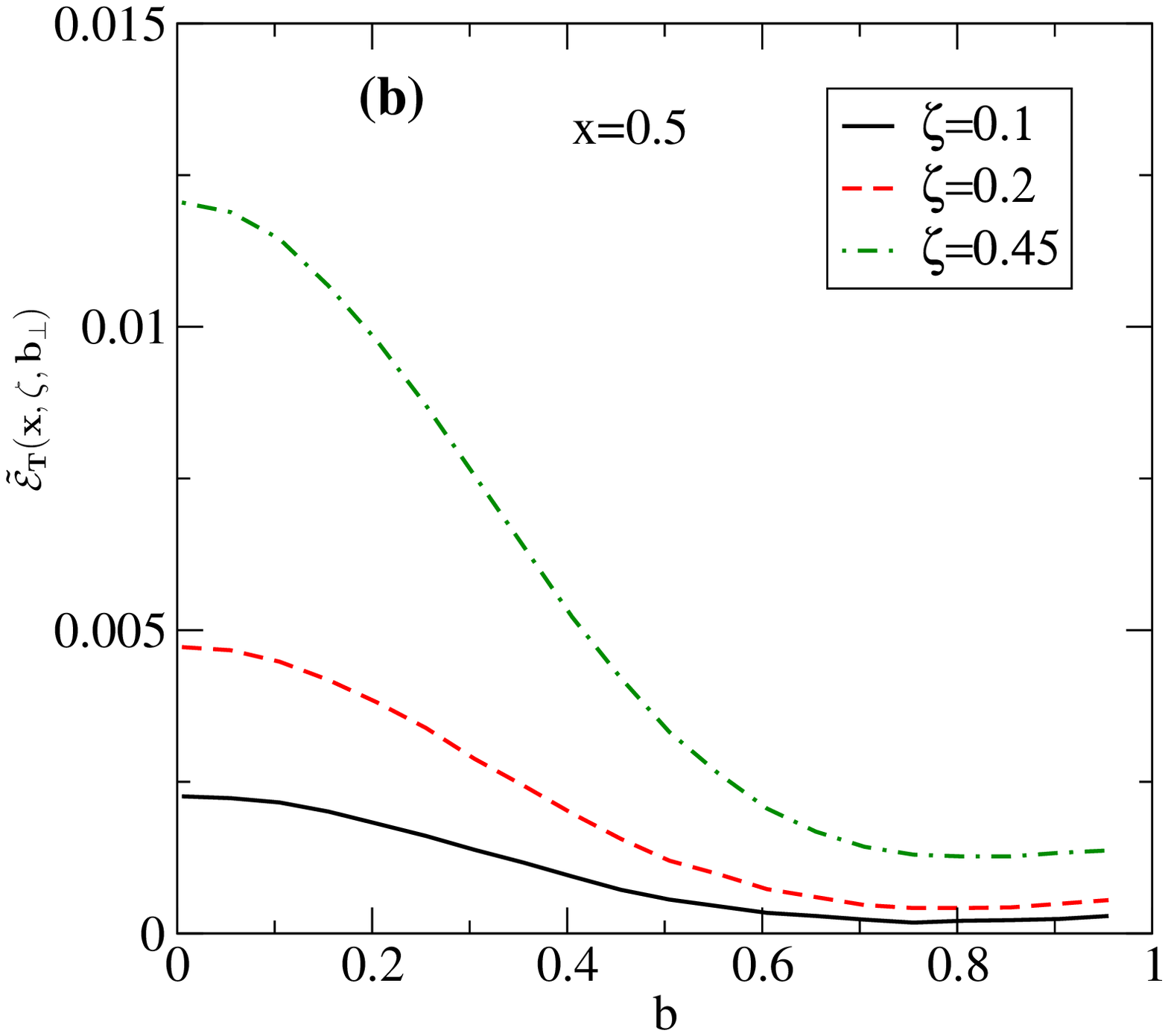}
\hspace{0.2cm}%
\includegraphics[width=7cm,height=7cm,clip]{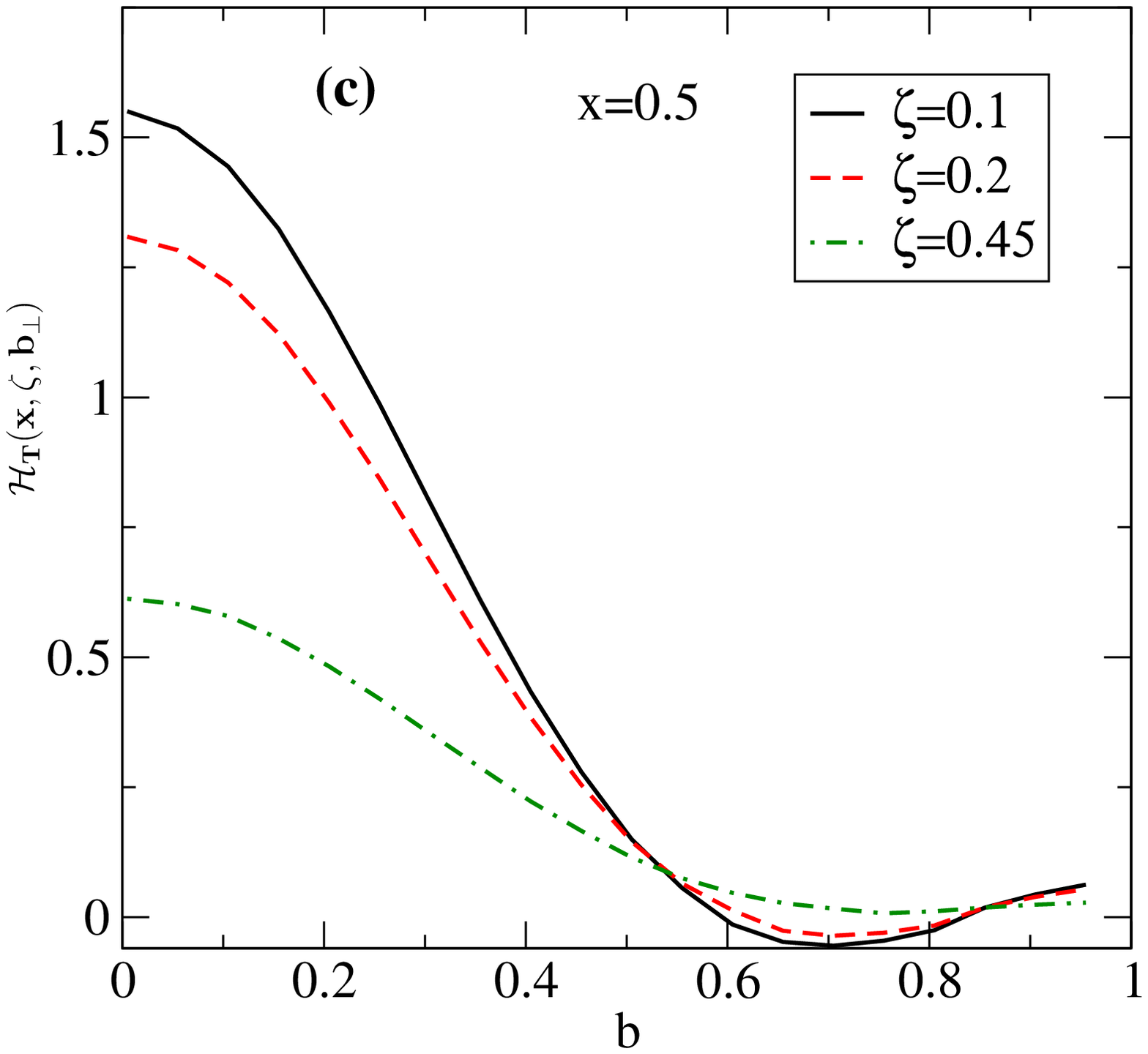}
\end{minipage}%
\caption{\label{fig4} (Color online) Fourier spectrum of the chiral-odd
GPDs vs. $\mid b_\perp \mid $  for fixed $x=0.5$  and
different values of $\zeta$} 
\end{figure}

Introducing the Fourier conjugate  $b_\perp$ (impact parameter) 
of the transverse momentum transfer $\Delta_\perp$, the GPDs can 
be expressed in impact parameter space. Like the chiral even counterparts,
chiral odd GPDs as well have interesting interpretation in impact parameter
space. The second moment of $H_T, E_T$ and ${\tilde H}_T$ is related to the
transverse component of the total angular momentum carried by transversely
polarized quarks in an unpolarized proton :
\be
<J^i>={s^i\over 4} \int dx x \Big [ H_T(x,0,0)+ 2 \tilde H_T (x,0,0)
+E_T (x,0,0) \Big ];
\ee
In the impact parameter 
space, at $\zeta=0$,  the second two terms denote a deformation in the
transversity asymmetry of quarks in an unpolarized target. This deformation
is due to the spin-orbit correlation of the constituents \cite{burchi, chiral}. 
This is similar
to the role played by the GPD $E(x,0,0)$ in Ji's sum rule for the longitudinal 
angular momentum. On the other hand, the combination $H_T-{t\over 2 M^2}
{\tilde H_T}$ in impact parameter space gives the correlation between the
transverse quark spin and the spin of the transversely polarized nucleon. 

\begin{figure}
\centering
\includegraphics[width=7cm,height=7cm,clip]{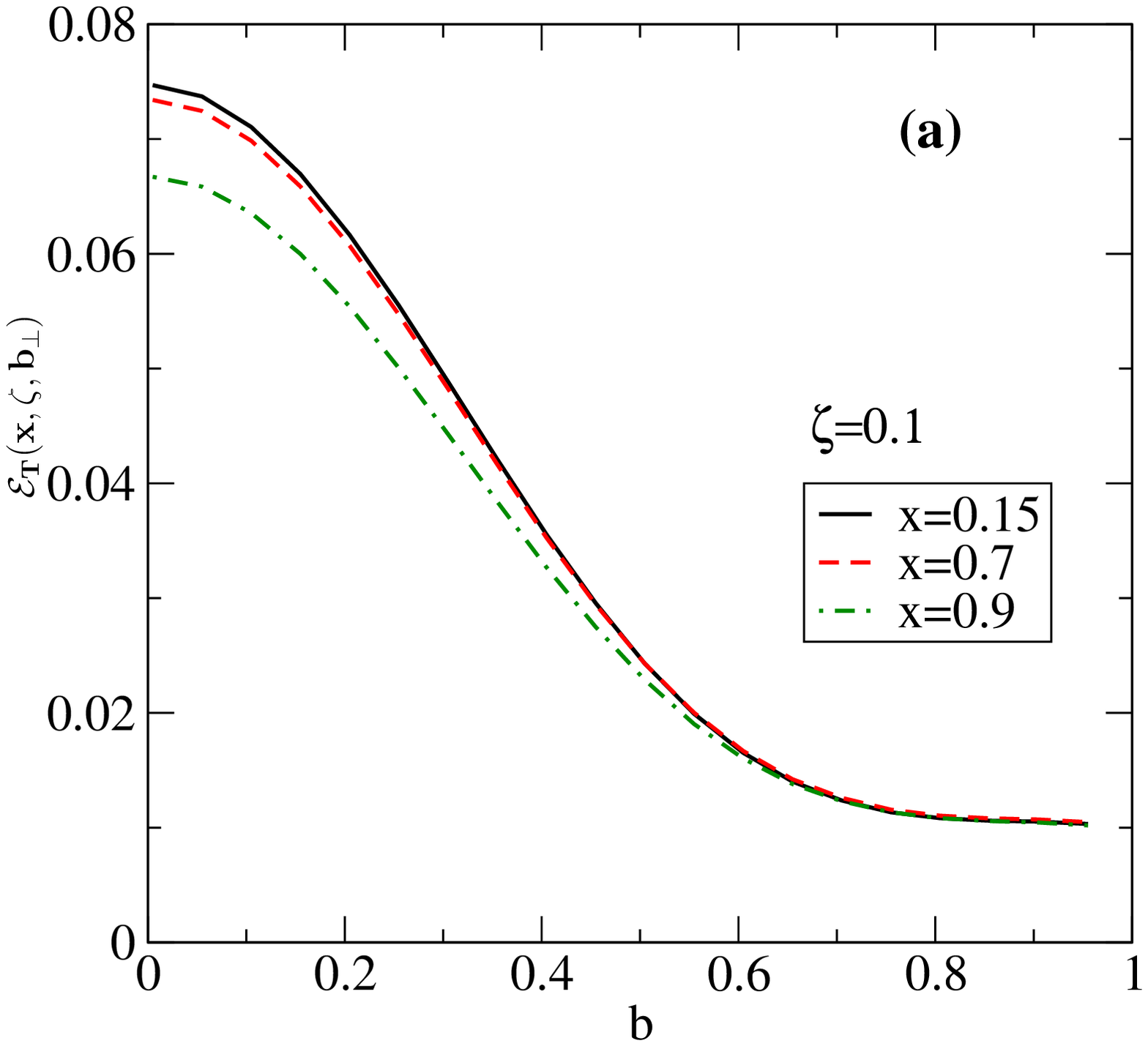}
\centering
\begin{minipage}[c]{0.9\textwidth}
\includegraphics[width=7cm,height=7cm,clip]{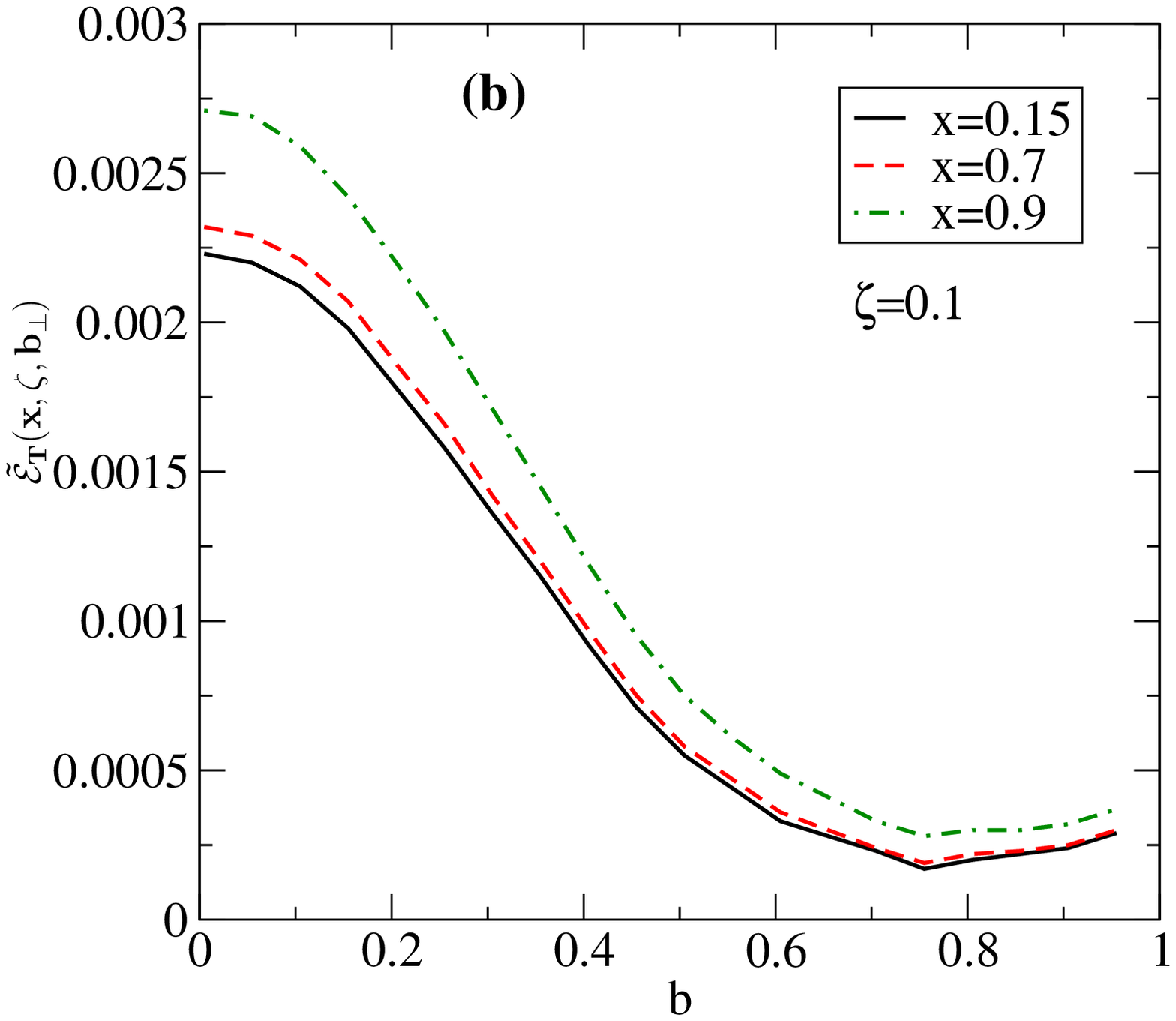}
\hspace{0.2cm}%
\includegraphics[width=7cm,height=7cm,clip]{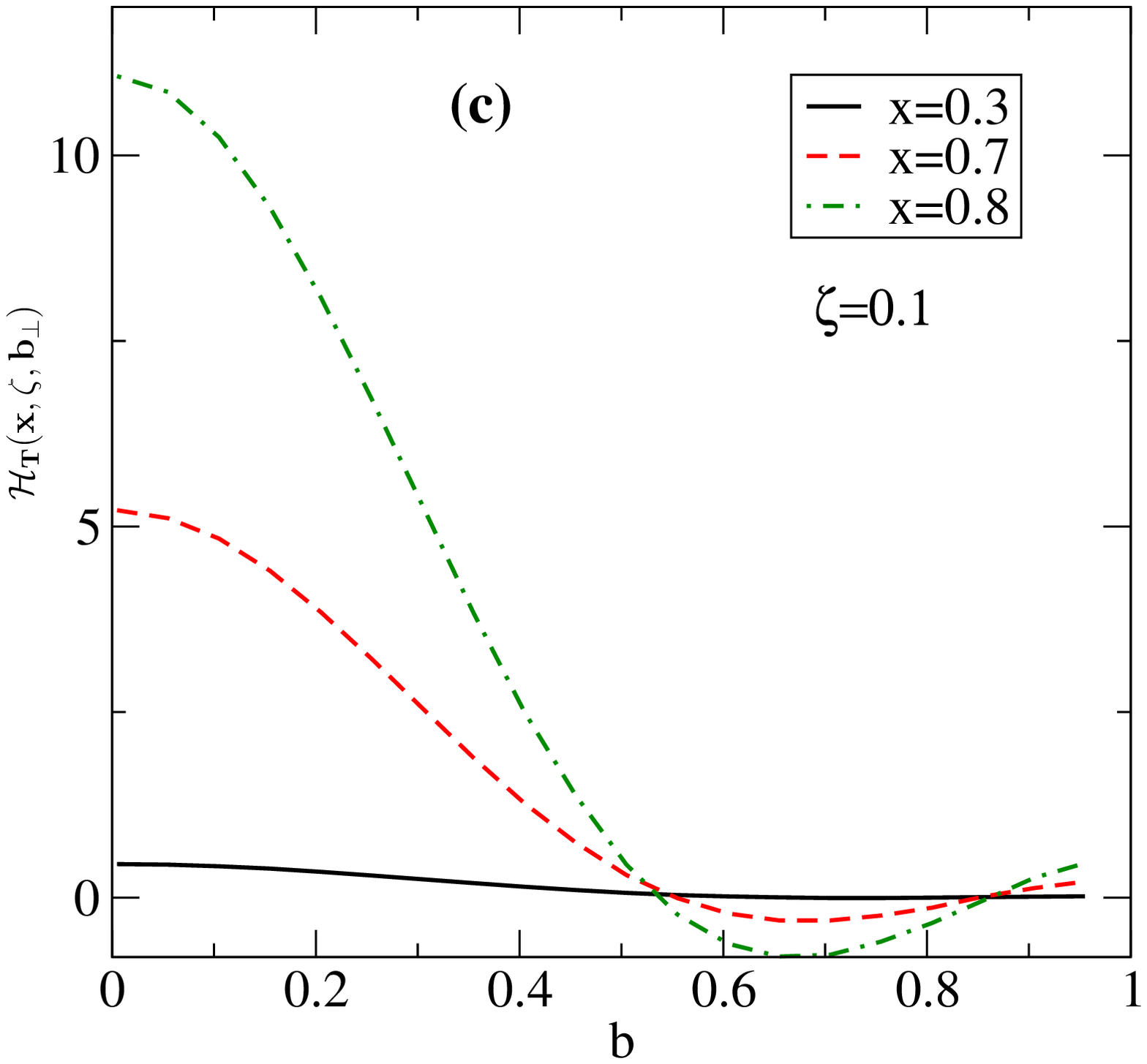}
\end{minipage}%
\caption{\label{fig5} (Color online) Fourier spectrum of the chiral-odd
GPDs vs. $\mid b_\perp \mid $  for fixed $\zeta$  and
different values of $x$} 
\end{figure}

The above picture was proposed in the limit of zero skewness $\zeta$ in
\cite{burchi, chiral}. In most experiments $\zeta$ is nonzero, and it is
of interest to investigate the chiral odd GPDs in $b_\perp$ space with
nonzero $\zeta$. The probability interpretation is no longer possible as now
the transverse position of the initial and final protons are different as
there is a finite momentum transfer in the longitudinal direction. The GPDs
in impact parameter space probe partons at transverse position $\mid b_\perp
\mid $ with the initial and final proton shifted by an amount of order $\zeta
\mid b_\perp \mid$. Note that this is independent of $x$ and and even when
GPDs are integrated over $x$ in an amplitude, this information is still
there \cite{markus2}. Thus the chiral odd GPDs in impact parameter space
gives the spin orbit correlations of partons in protons with their centers
shifted with respect to each other.      

Taking the Fourier transform with respect to the transverse momentum transfer
 $\Delta_\perp$ we get the GPDs in the transverse impact parameter space.
\be
{\cal E}_T(x,\zeta,b_\perp)&=&{1\over (2\pi)^2}\int d^2\Delta_\perp 
e^{-i\Delta_\perp \cdot b_\perp} E_T(x,\zeta,t)\nonumber \\
&=&{1\over 2 \pi}\int \Delta d\Delta J_0(\Delta b) E_T(x,\zeta,t),
\ee
where $\Delta=|\Delta_\perp|$ and $b=|b_\perp|$. The other impact 
parameter dependent GPDs $\tilde{\cal E}_T(x,\zeta,b_\perp)$ and 
${\cal H}_T(x,\zeta,b_\perp)$ can also be defined in the same way.

Fig.\ref{fig4}  shows the chiral odd  GPDs for nonzero $\zeta$ in impact
parameter space for different $\zeta$ and fixed $x=0.5$ as a function 
of $\mid b_\perp \mid$. As $\zeta$ increases the peak at $\mid b_\perp \mid
=0$ increases for $\tilde{\cal E}_T(x,\zeta,b_\perp)$  but 
decreases for ${\cal H}_T(x,\zeta,b_\perp)$ and 
${\cal E}_T(x,\zeta,b_\perp)$.
 It is to be noted that ${\cal H}_T(x,\zeta,b_\perp)$
for a free Dirac particle is expected to be a delta function; the smearing
in $\mid b_\perp \mid $ space is due to the spin correlation in the
two-particle LFWFs. 
 Fig. \ref{fig5} shows the plots of the above three functions
for fixed $\zeta$ and different values of $x$. For given $\zeta$, the peak
of ${\cal H}_T(x,\zeta,b_\perp)$ as well 
as $\tilde {\cal E}_T(x,\zeta,b_\perp)$
increases with increase of $x$, however for ${\cal E}_T(x,\zeta,b_\perp)$ 
it decreases.  

\begin{figure}
\centering
\includegraphics[width=7cm,height=7cm,clip]{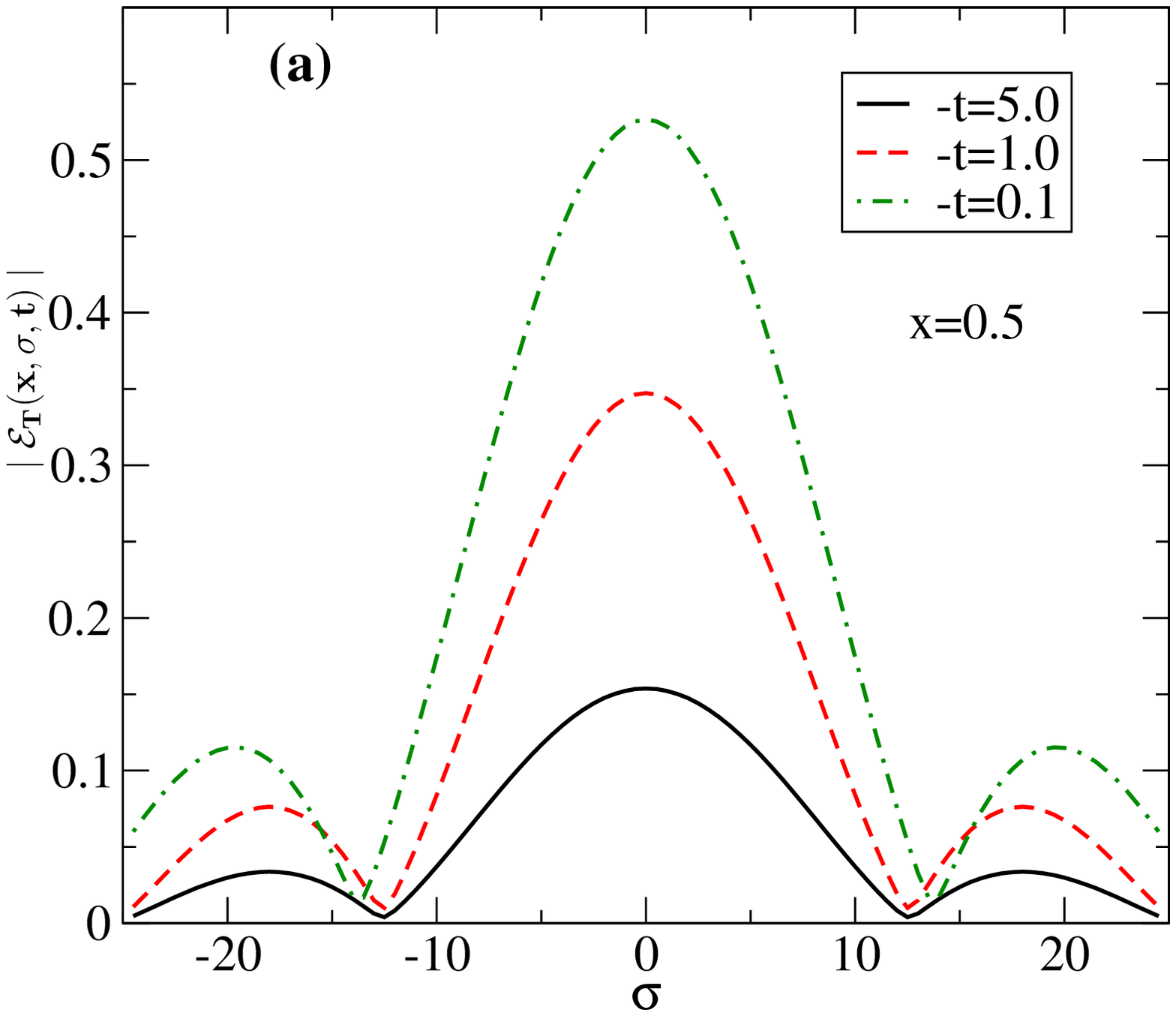}
\centering
\begin{minipage}[c]{0.9\textwidth}
\includegraphics[width=7cm,height=7cm,clip]{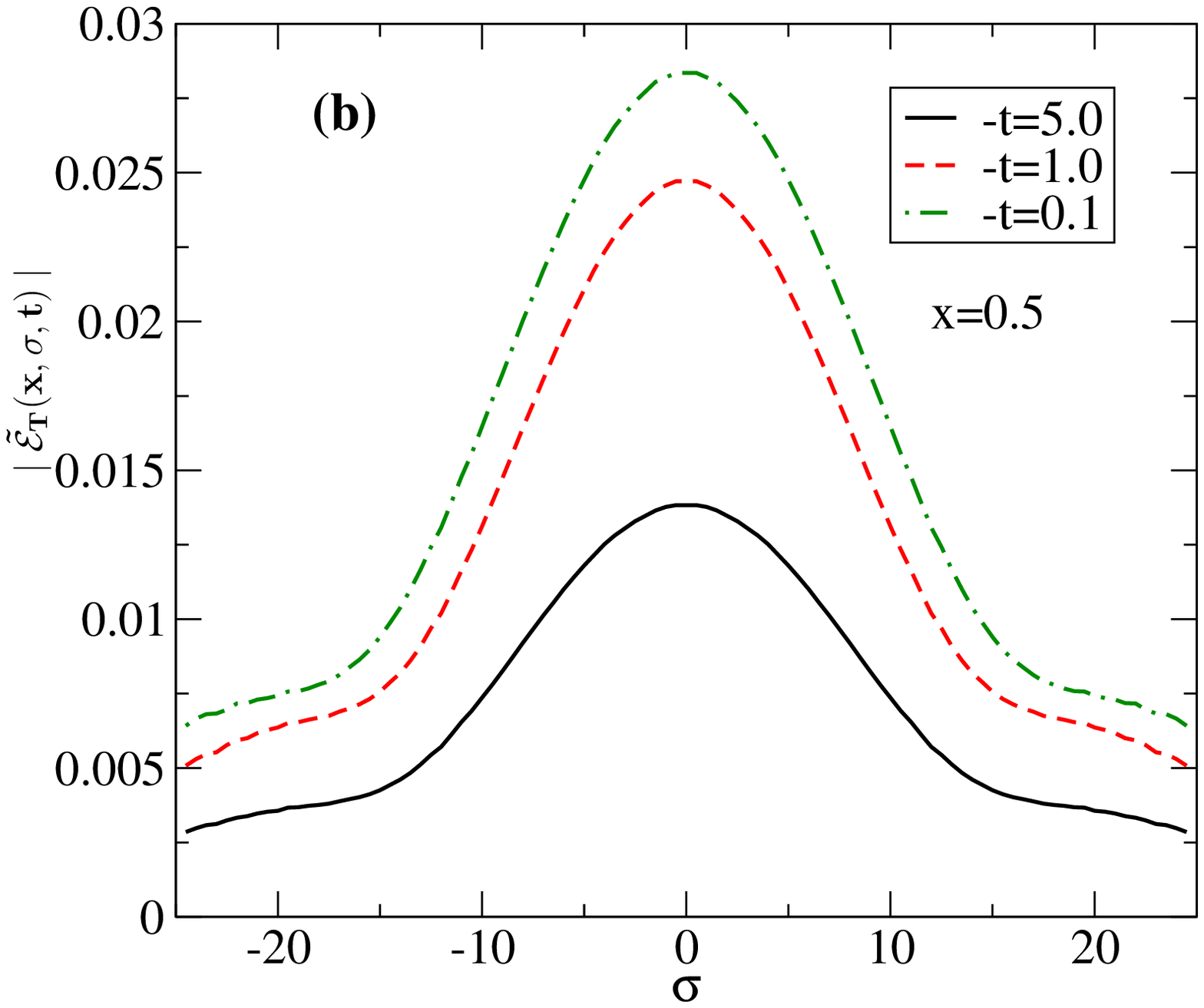}
\hspace{0.2cm}%
\includegraphics[width=7cm,height=7cm,clip]{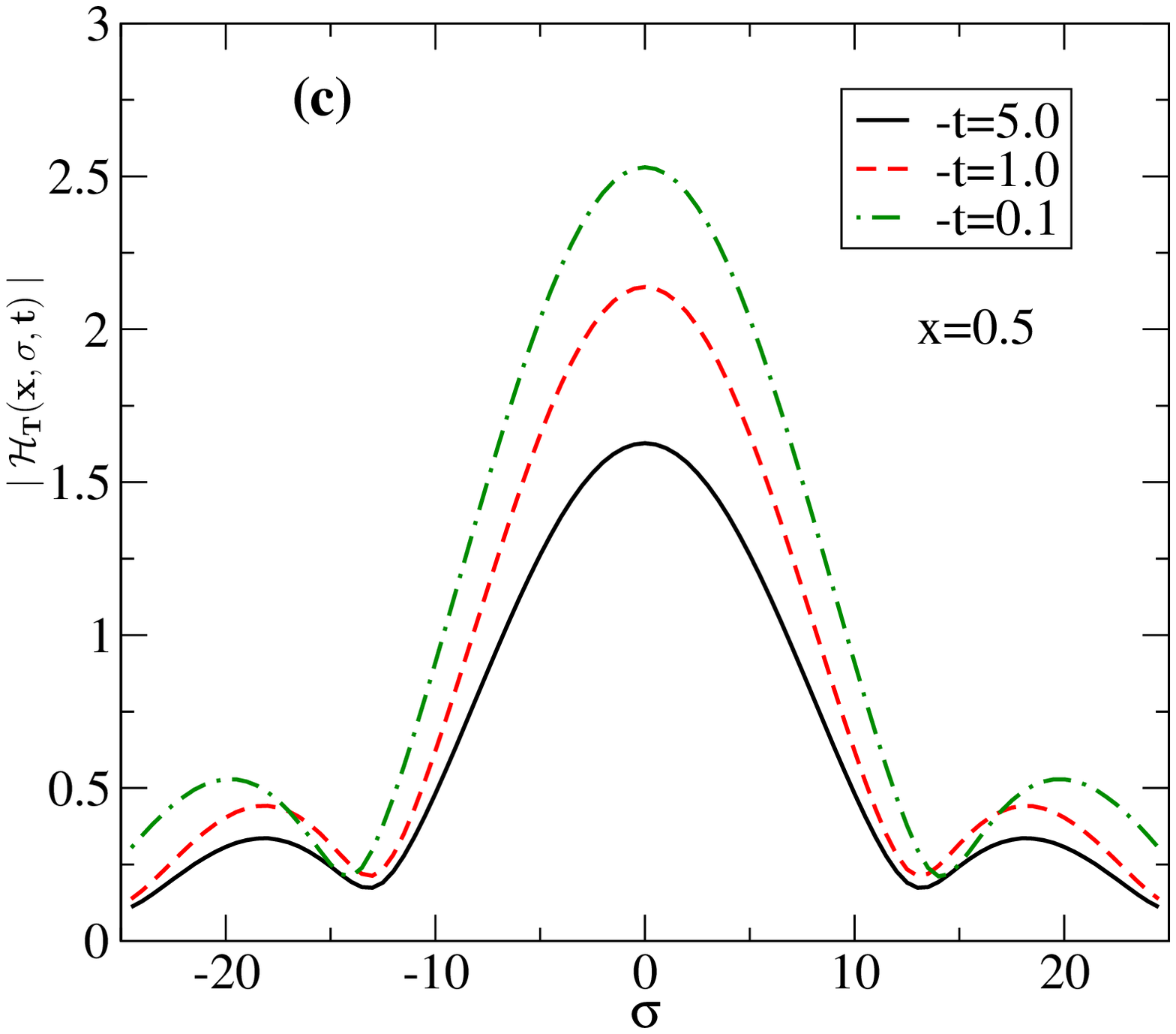}
\end{minipage}%
\caption{\label{fig6} (Color online) Fourier spectrum of the chiral-odd
GPDs vs. $\sigma$  for fixed $x$  and different values of $-t$ in
${\mathrm{MeV}^2}$.} 
\end{figure}

So far, we discussed about the chiral odd GPDs in transverse position space. 
In \cite{wigner}, a phase space distribution of quarks and gluons in the
proton is given in terms of the quantum mechanical Wigner distribution
$W(\vec{r}, \vec{p})$, in the rest frame of the proton, which are functions 
of three position and three
momentum coordinates. Wigner distributions are not accessible in experiment.
However, if one integrates two momentum components one gets a reduced Wigner
distribution $W_\Gamma(\vec{r},x)$ which is related to the GPDs by a Fourier
transform.  For given $x$, this gives a 3D position space picture of the
partons inside the proton. In the infinite momentum frame, rotational
symmetry is not there. Nevertheless, one can still define such a 3D position 
space distribution by taking a Fourier transform of the GPDs. Another point
is the dynamical effect of Lorentz boosts. If the probing wavelength is
comparable to or smaller than the Compton wavelength ${1\over M}$, where $M$
is the mass of the proton, electron-positron pairs will be created, as a
result, the static size of the system cannot be probed to a precision better
than  ${1\over M}$ in relativistic quantum theory. However, in light-front
theory, transverse boosts are Galilean boosts which do not involve dynamics.
So one can still express the GPDs in transverse position or impact parameter
space and this picture is not spoilt by relativistic corrections. However,
rotation involves dynamics here and rotational symmetry is lost. In
\cite{hadron_optics}, a longitudinal boost invariant impact parameter
$\sigma$ has been introduced which is conjugate to the
longitudinal momentum transfer $\zeta$. It was shown that the DVCS amplitude
expressed in terms of the variables $\sigma, b_\perp$ show diffraction
pattern analogous to diffractive scattering of a wave in optics where the
distribution in $\sigma$ measures the physical size of the scattering
center in a 1-D system. In analogy with optics, it was concluded that the
finite size of the $\zeta$ integration of the FT acts as a slit of finite
width and produces the diffraction pattern.

We define a boost invariant impact parameter conjugate to the  longitudinal 
momentum transfer  as 
$\sigma={1\over 2}b^-P^+$ \cite{hadron_optics}.  The chiral odd GPD $E_T$
in longitudinal position space is given by :  
\be
{\cal E}_T(x,\sigma,t)&=& {1\over 2\pi}\int_0^{\zeta_f}d\zeta 
e^{i{1\over 2}P^+\zeta b^-} E_T(x,\zeta,t)\nonumber \\
&=&{1\over 2 \pi}\int_0^{\zeta_f} d\zeta e^{i\sigma\zeta} E_T(x,\zeta,t).
\ee
Since we are concentrating only in the region $\zeta<x<1$, the upper 
limit of $\zeta$ integration $\zeta_f$ is given by $\zeta_{max}$ 
if $x$ is larger then $\zeta_{max}$, otherwise by $x$ if $x$ is smaller than
$\zeta_{max}$ where $\zeta_{max}$ is the maximum value of $\zeta$ allowed 
for a fixed $-t$: 
\be
\zeta_{max}={(-t)\over 2 M^2}\Big( \sqrt{1+{4 M^2\over (-t)}}-1 \Big).
\ee 
Similarly one can obtain ${\cal H}_T(x,\sigma,t)$ and $\tilde {\cal E}_T
(x,\sigma,t)$ as well. Fig. \ref{fig6} shows the plots of the Fourier spectrum of
 chiral odd GPDs in 
longitudinal position space  as a function of $\sigma$ for fixed $x=0.5$ and
different values of $t$. Both ${\cal E}_T(x,\sigma,t)$ and 
${\cal H}_T(x,\sigma,t)$ show diffraction pattern as observed for the DVCS
amplitude in \cite{hadron_optics}; the minima occur at the sames values of
$\sigma$ in both cases.  However $\tilde {\cal E}_T(x,\sigma,t)$
does not show diffraction pattern. 
This is due to the distinctively 
different behaviour of 
$\tilde E_T(x,\zeta,t)$ with $\zeta$ compared to that of  $E_T(x,\zeta,t)$ and
$H_T(x,\zeta,t)$. $\tilde E_T(x,\zeta,t)$ rises smoothly from zero and 
has no flat plateau in $\zeta$ and thus does not exhibit any diffraction
pattern when Fourier transformed with respect to $\zeta$. 
The position of first minima in  Fig.\ref{fig6} is determined by $\zeta_f$. 
For $-t=5.0$ and $1.0$, $\zeta_f \approx x=0.5$
and thus the first minimum appears at the same position while for  $-t=0.1$, 
$\zeta_f =\zeta_{max}\approx 0.45$ and the minimum appears  slightly shifted. 
This is analogous to the single slit optical diffraction pattern. 
$\zeta_f$ here plays the role of the slit width.
 Since the positions of the minima(measured from the centre of 
the diffraction pattern) are inversely proportional to the slit width, 
the minima 
move away from the centre as the slit width (i.e., $\zeta_f$) decreases.
The  optical analogy of the diffraction pattern in $\sigma$ space has been discussed in detail in 
\cite{hadron_optics} in the context of DVCS amplitudes.

\section{Conclusion}
In this work, we have studied the chiral-odd GPDs in transverse and
longitudinal position space. Working in light-front gauge, we presented 
overlap formulas for the chiral odd GPDs in terms of proton 
light-front wave functions both in the DGLAP and ERBL regions. In the first
case there is parton number conserving $n \to n$ overlap whereas in the
latter case, parton number changes by two in $n+1 \to n-1$ overlap.
We investigated them in the DGLAP region, when the skewness $\zeta$ is
less than $x$. We used a self consistent relativistic two-body model,
 namely the quantum
fluctuation  of an electron at one loop in QED. We used its most general 
form \cite{drell}, where we have a different mass for the external electron
and different masses for the internal electron and photon. 
The impact parameter space representations are obtained by taking Fourier transform of
the GPDs with respect to the transverse momentum transfer. It is known that 
\cite{chiral,burchi} the chiral odd GPDs provide important information on
the spin-orbit correlations of the transversely polarized partons in an
unpolarized nucleon, as well as the correlations between the transverse 
quark spin and the nucleon spin in the transverse polarized nucleon. When
$\zeta$ is non-zero, the initial and final proton are displaced in the
impact parameter space relative to each other by an amount proportional to
$\zeta$. As this is the region probed by most experiments, it is of interest
to investigate this. By taking a Fourier transform with respect to $\zeta$ we
presented the GPDs in the boost invariant longitudinal position space
variable
$\sigma$. $H_T$ and $E_T$ show diffraction pattern in $\sigma$ space.
Further work is needed to investigate this behaviour and to study its model
dependence.
\section{acknowledgment}
The work of DC is supported by Marie-Curie (IIF) Fellowship. AM thanks 
DST Fasttrack scheme, Govt. of India  for financial support
for completing this work.

\end{document}